\documentclass[10pt,twocolumn,english,manuscript]{revtex4}
\usepackage[T1]{fontenc}
\usepackage[latin9]{inputenc}
\usepackage{color}
\usepackage{amsmath}
\usepackage{amssymb}
\usepackage{graphicx}
\usepackage{wasysym}

\makeatletter
\@ifundefined{textcolor}{}
{%
 \definecolor{BLACK}{gray}{0}
 \definecolor{WHITE}{gray}{1}
 \definecolor{RED}{rgb}{1,0,0}
 \definecolor{GREEN}{rgb}{0,1,0}
 \definecolor{BLUE}{rgb}{0,0,1}
 \definecolor{CYAN}{cmyk}{1,0,0,0}
 \definecolor{MAGENTA}{cmyk}{0,1,0,0}
 \definecolor{YELLOW}{cmyk}{0,0,1,0}
}

\usepackage{babel}

\makeatother

\usepackage{babel}
\begin{document}

\title{Tuning the thermal entanglement in a Ising-$XXZ$ diamond chain with
two impurities}

\author{I. M. Carvalho$^{1}$, O. Rojas$^{1,2}$, S. M. de Souza$^{1}$ and
M. Rojas$^{1}$}

\affiliation{$^{1}$Departamento de Física, Universidade Federal de Lavras, 37200-000,
Lavras-MG, Brasil}

\affiliation{$^{2}$ICTP, Strada Costiera 11, I-34151 Trieste, Italy}
\begin{abstract}
We study the local thermal entanglement in \textcolor{black}{a spin-1/2}
Ising-$XXZ$ diamond \textcolor{black}{structure} with two impurities.
In this spin chain, we have the two impurities with an isolated $XXZ$
dimer between them. We focus on the study of the thermal entanglement
in this dimer. The main goal of this paper is to provide a good understanding
of the effect of impurities in the entanglement of the model. This
model is exactly solved by a rigorous treatment based on the transfer-matrix
method. \textcolor{black}{Our results show that the entanglement can
be tuned by varying the impurities parameters}\textcolor{red}{{} }\textcolor{black}{in
this system. In addition,} it is shown that the thermal entanglement
for such a model exhibits a clear \textcolor{black}{performance improvement
when} we control and manipulate \textcolor{black}{the impurities}
compared to the original model without impurities. \textcolor{black}{Finally,
the impurities can be} manipulated \textcolor{black}{to locally}\textcolor{red}{{}
}\textcolor{black}{control the} thermal entanglement, unlike the original
model where it is done globally.
\end{abstract}
\maketitle

\section{Introduction}

Entanglement is one the most fascinating feature of quantum mechanics
due to its nonclassical feature and potential applications in quantum
information processing \cite{bene}. The spin chains with Heisenbeg
interaction are regarded as \textcolor{black}{one} natural candidate
in order to be used in the field of quantum information processing
\cite{loss}.\textcolor{red}{{} }\textcolor{black}{In recent years,
numerous studies have been performed on the thermal entanglement in
various Heisenberg models} \cite{kam,zhang,bose,wang}.

More recently, considerable attention has been \textcolor{black}{given
to the} rigorous treatment of various versions of the Ising-Heisenberg
diamond chains \cite{cano,lis}. The exactly solvable Ising-Heisenbeg
spin models provide a reasonable quantitative description of the thermal
entanglement behavior. The thermal entanglement properties of several
kinds of Ising-Heisenberg diamond \textcolor{black}{chain}\textcolor{red}{{}
}models have been studied in Ising-$XXZ$ model on diamond chain \cite{moi},
Ising-$XYZ$ model on diamond chain \cite{rojas} and the mixed spin-1/2
and spin-1 Ising-Heisenberg model on diamond chain \cite{ana}. The
quantum teleportation via a couple of quantum channels composed of
$XXZ$ dimers in an Ising-$XXZ$ diamond chain has been reported \cite{rojas-1}.
Furthermore, the quantum correlation has been quantified by the trace
distance discord to describe the quantum critical behaviors in the
Ising-$XXZ$ diamond structure at finite temperature \cite{cheng}.

The impurity plays an important role in solid state physics \cite{stol,fal}.
In particular, the spin chains with impurities can be viewed as a
spin chain with \textcolor{black}{impurity at the} spin site \cite{xi},
and the strength of interactions between \textcolor{black}{the spin
of}\textcolor{red}{{} }impurity and its neighboring spins may be different
from that between normal spins. The impurity effects of transverse
Ising model with multi-impurity has been extensively studied \cite{hu,sun}.
On the other hand, impurity effects on quantum entanglement have been
consider in Heisenberg spin chain \cite{osenda,fu}. More recently
the quantum discord based in Heisenberg chain with impurities \cite{ji}
has also been studied. However, the problem of the thermal entanglemet
in the Ising-Heisenberg chains with impurities has not been studied.

In this paper we provide a new approach to control the thermal entanglement
in Ising-Heisenberg-type spin chains by introducing one or a few impurities
into the structure of the model. The way to control the entanglement
is done locally, is to say, modulating the interaction parameters
of the Heisenberg spins dimers and/or the interaction constants of
the Ising spins \textcolor{black}{in} the impurities. It should be
noted that the addition of impurities allows us to have more robust
entanglement -choosing the appropriate parameters- than any Ising-Heisenberg
chain without impurities. In particular, we show these results in
the case of an Ising-$XXZ$ diamond chain with two impurities inserted
in it, in this configuration we consider a $XXZ$ dimer between \textcolor{black}{the}\textcolor{red}{{}
}two impurities and we will focus on the analysis of the thermal entanglement
in this dimer. We demonstrate that entanglement \textcolor{black}{can}
be effectively controlled through the impurities \textcolor{black}{of}
the model, and can be enhanced or weakened \textcolor{black}{for}
some suitable parameters.

The paper is structured as follows. In Sec. II we present the Ising-$XXZ$
model with two impurity. In Sec. III, we obtain the exact solution
of the model via the transfer-matrix approach, which allows a straightforward
calculation of the average reduced density operator of the isolated
$XXZ$ dimer between two impurities. In Sec. IV, we discuss the thermal
entanglement of the Heisenberg reduced density operator of the model,
such as concurrence and threshold temperature. Finally, concluding
remarks are given in Sec. V.

\section{Ising-$XXZ$ diamond chain with two impurities}

\begin{figure}
\includegraphics[scale=0.45]{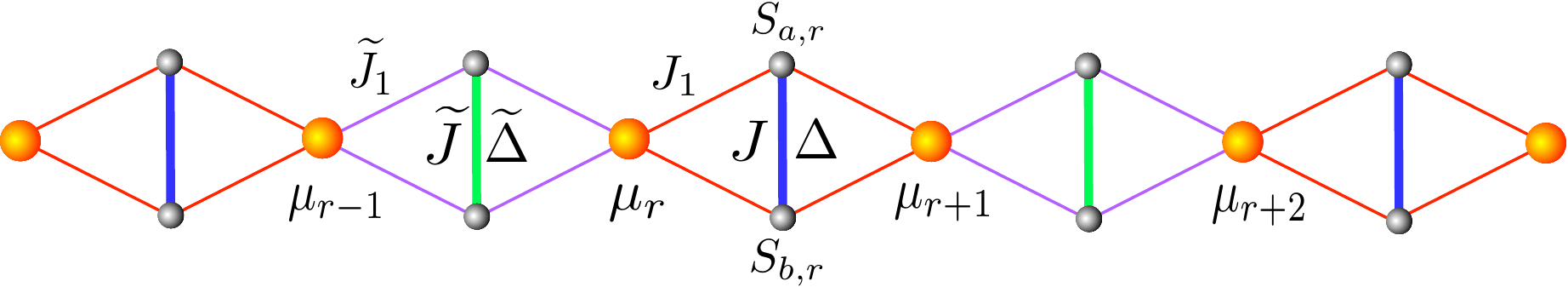}\caption{\label{fig:diamond}(Color online) A fragment from the Ising-$XXZ$
diamond chain with two impurities. The Ising spins is denoted by $\mu_{r}$,
while $S_{a,r}$ are the Heisenberg spins.}
\end{figure}

Let us consider the Ising-$XXZ$ model with two impurities on a diamond
chain in the presence of an external magnetic field. The various interaction
parameters $\left(J,\Delta,J_{1}\right)$ are homogeneous along the
diamond chain, except in the impurities where the corresponding parameters
$\left(\widetilde{J},\widetilde{\Delta},\widetilde{J}_{1}\right)$
\textcolor{black}{are} different as shown in the Fig. \ref{fig:diamond}.
The Ising-$XXZ$ model consist of the Ising spins $\mu_{i}$ locate
at the nodal lattice sites as well as, interstitial Heisenberg spins
$S_{a(b)}^{\alpha}(\alpha=x,y,z)$. 

Thus, the total hamiltonian for the Ising-$XXZ$ diamond chain with
two impurities embedded in the model is given by $\mathcal{H}=\sum_{i=1}^{N}\mathcal{H}_{i}$,
where

\[
\mathcal{H}_{i}=\mathcal{H}_{i}^{host}+\mathcal{H}_{i}^{imp}+\mathcal{H}_{i}^{iso},
\]
and

\[
\begin{array}{cl}
\mathcal{H}_{i}^{host}= & J\left(\mathbf{S}_{a,i},\mathbf{S}_{b,i}\right)_{\Delta}+J_{1}\left(S_{a,i}^{z}+S_{b,i}^{z}\right)\left(\mu_{i}+\mu_{i+1}\right)\\
 & -h\left(S_{a,i}^{z}+S_{b,i}^{z}\right)-\frac{h}{2}\left(\mu_{i}+\mu_{i+1}\right),\\
 & \mathrm{for}\:i=1,2,\ldots,r-2,r+2,\ldots,N
\end{array}
\]
is the hamiltonian of the host chain with $(\mathbf{S}_{a,i},\:\mathbf{S}_{b,i})_{\Delta}=(S_{a,i}^{x}S_{b,i}^{x}+S_{a,i}^{y}S_{b,i}^{y}+\Delta S_{a,i}^{z}S_{b,i}^{z})$
and $\mu_{i}=\pm1/2$. 

Whereas

\[
\begin{array}{cl}
\mathcal{H}_{i}^{imp}= & \widetilde{J}\left(\mathbf{S}_{a,i},\mathbf{S}_{b,i}\right)_{\widetilde{\Delta}}+\widetilde{J}_{1}\left(S_{a,i}^{z}+S_{b,i}^{z}\right)\left(\mu_{i}+\mu_{i+1}\right)\\
 & -h\left(S_{a,i}^{z}+S_{b,i}^{z}\right)-\frac{h}{2}\left(\mu_{i}+\mu_{i+1}\right),\\
 & \mathrm{for}\:i=r-1,r+1
\end{array}
\]
is the hamiltonian of the impurities of the model. Finally,

\[
\begin{array}{cl}
\mathcal{H}_{i}^{iso}= & J\left(\mathbf{S}_{a,i},\mathbf{S}_{b,i}\right)_{\Delta}+J_{1}\left(S_{a,i}^{z}+S_{b,i}^{z}\right)\left(\mu_{i}+\mu_{i+1}\right)\\
 & -h\left(S_{a,i}^{z}+S_{b,i}^{z}\right)-\frac{h}{2}\left(\mu_{i}+\mu_{i+1}\right)\\
 & \mathrm{for}\:i=r
\end{array}
\]
is the hamiltonian of the plaquette isolated in between of the impurities.

Here, $J$ and $\Delta$ denote the $XXZ$ interaction within the
interstitial Heisenberg dimers, while $J_{1}$ label the interactions
between the nodal Ising spins and interstitial Heisenberg spins, and
the quantities $\widetilde{J}=J\left(1+\alpha\right)$, $\widetilde{\Delta}=\Delta\left(1+\gamma\right)$
and $\widetilde{J}_{1}=J_{1}\left(1+\eta\right)$ are defined analogously
for the impurities. In addition $\alpha$, $\gamma$ and $\eta$ measures
the strength of the impurities. Finally, $h$ is the magnetic field
along the $z$-axis. 

\textcolor{black}{After straightforward calculation, the eigenvalues
for the host hamiltonian}\textcolor{red}{{} }\textcolor{black}{$\mathcal{H}_{i}^{host}$
can be written as}\textcolor{red}{{} }
\begin{align*}
\mathcal{\varepsilon}_{i1,i4}= & \frac{J\Delta}{4}\pm\left(J_{1}\mp\frac{h}{2}\right)(\mu_{i}+\mu_{i+1})\mp\frac{h}{2},\\
\mathcal{\varepsilon}_{i2,i3}= & -\frac{J\Delta}{4}\pm\frac{J}{2}-\frac{h}{2}(\mu_{i}+\mu_{i+1}),
\end{align*}
where their corresponding eigenstate in terms of standard basis $\{|00\rangle,|01\rangle,|10\rangle,|11\rangle\}$
are given respectively by 
\begin{align}
|\varphi_{i1}\rangle= & |00\rangle_{i},\\
|\varphi_{i2}\rangle= & \frac{1}{\sqrt{2}}\left(|01\rangle_{i}+|10\rangle_{i}\right),\label{eq:phi2}\\
|\varphi_{i3}\rangle= & \frac{1}{\sqrt{2}}\left(|01\rangle_{i}-|10\rangle_{i}\right),\\
|\varphi_{i4}\rangle= & |11\rangle_{i}.
\end{align}

On the other hand, the eigenvalues of the impurity hamiltonian $\mathcal{H}_{i}^{imp}$
is given by

\[
\begin{array}{cl}
\mathcal{\epsilon}_{i1,i4}= & \frac{\widetilde{J}\widetilde{\Delta}}{4}\pm\left(\widetilde{J}_{1}\mp\frac{h}{2}\right)(\mu_{i}+\mu_{i+1})\mp\frac{h}{2},\\
\mathcal{\epsilon}_{i2,i3}= & -\frac{\widetilde{J}\widetilde{\Delta}}{4}\pm\frac{\widetilde{J}}{2}-\frac{h}{2}(\mu_{i}+\mu_{i+1}),
\end{array}
\]
with corresponding eigenvectors

\[
\begin{array}{cl}
|\widetilde{\varphi}_{i1}\rangle= & |00\rangle_{i},\\
|\widetilde{\varphi}_{i2}\rangle= & \frac{1}{\sqrt{2}}\left(|01\rangle_{i}+|10\rangle_{i}\right),\\
|\widetilde{\varphi}_{i3}\rangle= & \frac{1}{\sqrt{2}}\left(|01\rangle_{i}+|10\rangle_{i}\right),\\
|\widetilde{\varphi}_{i4}\rangle= & |11\rangle_{i},
\end{array}
\]
here $i=r-1$ and $i=r+1$. \textcolor{black}{The spectrum shown above
allows us to construct the partition function of the Ising-Heisenberg
spin chain.}\textcolor{red}{{} }

\section{The partition function and density operator}

In order to measure the thermal entanglement, we first need to obtain
a partition function on a diamond chain. \textcolor{black}{Our} model
can be solved exactly using decoration transformation and transfer-matrix
approach \cite{baxter}. In this approach we will define the following
operator as a function of Ising spin particles $\mu_{i}$ and $\mu_{i+1}$,

\begin{equation}
\varrho(\mu_{i},\mu_{i+1})=\sum_{j=1}^{4}\mathrm{e}^{-\beta\varepsilon_{ij}(\mu_{i},\mu_{i+1})}|\varphi_{ij}\rangle\langle\varphi_{ij}|,
\end{equation}
where $\beta=1/k_{B}T$ with the Boltmann's constant $k_{B}$ and
$T$ is the absolute temperature.

Straightforwardly, we can obtain the Boltzmann factor by tracing out
over the two-qubit operator, 
\begin{equation}
w(\mu_{i},\mu_{i+1})=\mathrm{tr}_{ab}\left(\varrho(\mu_{i},\mu_{i+1})\right)=\sum_{j=1}^{4}\mathrm{e}^{-\beta\varepsilon_{ij}(\mu_{i},\mu_{i+1})},\label{eq:w-def}
\end{equation}
and Boltzmann factor for impurity is given by

\[
\widetilde{w}(\mu_{i},\mu_{i+1})=\sum_{j=1}^{4}\mathrm{e}^{-\beta\epsilon_{ij}(\mu_{i},\mu_{i+1})},
\]
where $i=r-1$ and $i=r+1$.

The canonical partition function of the Ising-$XXZ$ diamond chain
with two impurities can be written in terms of Boltzmann factors through
the relation

\begin{align}
Z_{N}= & \sum_{\{\mu\}}w(\mu_{1},\mu_{2})\ldots\widetilde{w}(\mu_{r-1},\mu_{r})w(\mu_{r},\mu_{r+1})\times\nonumber \\
 & \widetilde{w}(\mu_{r+1},\mu_{r+2})\ldots w(\mu_{N},\mu_{1}).\label{eq:rho-df-1}
\end{align}

Using the transfer-matrix notation, we can write the partition function
of the diamond chain straightforwardly by $Z_{N}=\mathrm{tr}\left(\widetilde{W}W\widetilde{W}W^{N-3}\right)$,
where the transfer-matrix takes the following form 
\begin{equation}
W=\left[\begin{array}{cc}
w(\frac{1}{2},\frac{1}{2}) & w(\frac{1}{2},-\frac{1}{2})\\
w(-\frac{1}{2},\frac{1}{2}) & w(-\frac{1}{2},-\frac{1}{2})
\end{array}\right].\label{eq:W}
\end{equation}
Similarly for impurities, we have

\[
\widetilde{W}=\left[\begin{array}{cc}
\widetilde{w}(\frac{1}{2},\frac{1}{2}) & \widetilde{w}(\frac{1}{2},-\frac{1}{2})\\
\widetilde{w}(-\frac{1}{2},\frac{1}{2}) & \widetilde{w}(-\frac{1}{2},-\frac{1}{2})
\end{array}\right].
\]
The transfer matrix elements are denoted by $w_{++}\equiv w(\frac{1}{2},\frac{1}{2})$,
$w_{+-}\equiv w(\frac{1}{2},-\frac{1}{2})$, $w_{--}\equiv w(-\frac{1}{2},-\frac{1}{2})$
and $\widetilde{w}_{++}\equiv\widetilde{w}(\frac{1}{2},\frac{1}{2})$,
$\widetilde{w}_{+-}\equiv\widetilde{w}(\frac{1}{2},-\frac{1}{2})$,
$\widetilde{w}_{--}\equiv\widetilde{w}(-\frac{1}{2},-\frac{1}{2})$.

The diagonalization of the transfer matrix (\ref{eq:W}), will provide
the followings eigenvalues 
\begin{equation}
\Lambda_{\pm}=\frac{w_{++}+w_{--}\pm Q}{2},
\end{equation}
assuming that $Q=\sqrt{\left(w_{++}-w_{--}\right)^{2}+4w_{+-}^{2}}$.
Therefore, the partition function for finite chain under periodic
boundary conditions is given by

\begin{equation}
Z_{N}=A\Lambda_{+}^{N-3}+D\Lambda_{-}^{N-3},
\end{equation}
where $A=a_{+}^{2}\Lambda_{+}+b_{+}b_{-}\Lambda_{-},$ $D=a_{-}^{2}\Lambda_{-}+b_{+}b_{-}\Lambda_{+}$
and 

\[
\begin{array}{cc}
a_{\pm}= & \frac{\pm4w_{+-}\widetilde{w}_{+-}\pm\left(w_{++}-w_{--}\right)\left(\widetilde{w}_{++}-\widetilde{w}_{--}\right)+Q\left(\widetilde{w}_{++}+\widetilde{w}_{--}\right)}{2Q}\\
b_{\pm}= & \frac{\left(w_{++}-w_{--}\pm Q\right)\left[\mp\left(\widetilde{w}_{++}-\widetilde{w}_{--}\right)\pm\widetilde{w}_{+-}\left(w_{++}-w_{--}\right)\right]}{2Qw_{+-}}.
\end{array}
\]

In the thermodynamic limit ($N\rightarrow\infty$) the partition function
will be simplified, which results in 
\[
Z_{N}=A\Lambda_{+}^{N-3}.
\]

\subsection{Average reduced density operator }

In order to calculate the average reduced density operator $\rho$,
we will use the two-qubit Heisenberg operator $\varrho$, the matrix
representation of this two-qubit operator is given by 
\begin{equation}
\varrho(\mu_{i},\mu_{i+1})=\left[\begin{array}{cccc}
\varrho_{1,1} & 0 & 0 & 0\\
0 & \varrho_{2,2} & \varrho_{2,3} & 0\\
0 & \varrho_{3,2} & \varrho_{3,3} & 0\\
0 & 0 & 0 & \varrho_{4,4}
\end{array}\right],
\end{equation}
where
\begin{align}
\varrho_{1,1}(\mu_{i},\mu_{i+1})= & \mathrm{e}^{-\beta\varepsilon_{i1}(\mu_{i},\mu_{i+1})},\nonumber \\
\varrho_{2,2}(\mu_{i},\mu_{i+1})= & \frac{1}{2}\left(\mathrm{e}^{-\beta\varepsilon_{i2}(\mu_{i},\mu_{i+1})}+\mathrm{e}^{-\beta\varepsilon_{i3}(\mu_{1},\mu_{i+1})}\right),\nonumber \\
\varrho_{2,3}(\mu_{i},\mu_{i+1})= & \frac{1}{2}\left(\mathrm{e}^{-\beta\varepsilon_{i2}(\mu_{i},\mu_{i+1})}-\mathrm{e}^{-\beta\varepsilon_{i3}(\mu_{i},\mu_{i+1})}\right),\nonumber \\
\varrho_{4,4}(\mu_{i},\mu_{i+1})= & \mathrm{e}^{-\beta\varepsilon_{i4}(\mu_{i},\mu_{i+1})}.
\end{align}

In order to investigate the thermal entanglement for the system, we
will perform the reduced density operator bounded by Ising particles
along a diamond chain. In particularly, we focus in the average reduced
density operator for the isolated two-qubit Heisenberg located between
the impurities. The reduced density matrix elements can be written
as 
\begin{align}
\rho_{k,l}= & \frac{1}{Z_{N}}\sum_{\{\mu\}}w(\mu_{1},\mu_{2})\ldots\widetilde{w}(\mu_{r-1},\mu_{r})\varrho_{k,l}(\mu_{r},\mu_{r+1})\times\nonumber \\
 & \widetilde{w}(\mu_{r+1},\mu_{r+2})\ldots w(\mu_{N},\mu_{1}).\label{eq:rho-df}
\end{align}

Using the transfer-matrix notation, the elements can be alternatively
rewritten as

\[
\begin{array}{rl}
\rho_{k,l}= & \frac{1}{Z_{N}}tr\left(W^{r-2}\widetilde{W}P_{k,l}\widetilde{W}W^{N-r-1}\right)\\
= & \frac{1}{Z_{N}}tr\left(\widetilde{W}P_{k,l}\widetilde{W}W^{N-3}\right),
\end{array}
\]
where

\begin{equation}
P_{k,l}=\left[\begin{array}{cc}
\varrho_{k,l}(++) & \varrho_{k,l}(+,-)\\
\varrho_{k,l}(-,+) & \varrho_{k,l}(-,-)
\end{array}\right],
\end{equation}
and $\varrho_{k,l}(++)\equiv\varrho_{k,l}(\tfrac{1}{2},\tfrac{1}{2})$,
$\varrho_{k,l}(+-)\equiv\varrho_{k,l}(\tfrac{1}{2},-\tfrac{1}{2})$,
$\varrho_{k,l}(-+)\equiv\varrho_{k,l}(-\tfrac{1}{2},\tfrac{1}{2})$,
$\varrho_{k,l}(--)\equiv\varrho_{k,l}(-\tfrac{1}{2},-\tfrac{1}{2})$.

The corresponding matrix $U$ that diagonalizes the transfer matrix
$W$ can be given by 
\begin{align}
U=\left[\begin{array}{cc}
\Lambda_{+}-w_{--} & \Lambda_{-}-w_{--}\\
w_{+-} & w_{+-}
\end{array}\right],
\end{align}
and 
\begin{align}
U^{-1}=\left[\begin{array}{cc}
\frac{1}{Q} & -\frac{\Lambda_{-}-w_{--}}{Qw_{+-}}\\
-\frac{1}{Q} & \frac{\Lambda_{+}-w_{--}}{Qw_{+-}}
\end{array}\right].
\end{align}

Finally, the individual matrix elements of the averaged reduced density
operator for the isolated dimer between two impurities defined in
Eq.(\ref{eq:rho-df}) must be expressed by

\begin{equation}
\rho_{k,l}=\tfrac{\mathrm{tr}\left(U^{-1}\widetilde{W}P_{k,l}\widetilde{W}U\left[\begin{smallmatrix}\Lambda_{+}^{N-3} & 0\\
0 & \Lambda_{-}^{N-3}
\end{smallmatrix}\right]\right)}{A\Lambda_{+}^{N-3}+D\Lambda_{-}^{N-3}}.
\end{equation}
This result is valid for arbitrary number $N$ of cells in a diamond
chain.

In the thermodynamic limit ($N\rightarrow\infty$) and assuming $\left(\Lambda_{-}/\Lambda_{+}\right)^{N}\rightarrow0$
in this limit, the reduced density operator elements after a cumbersome
algebra, is given by

\begin{align*}
\rho_{k,l}= & \frac{1}{A}\left\{ a_{+}^{2}\mathcal{M}_{k,l}^{+}+a_{+}b_{+}\mathcal{N}_{k,l}^{-}+a_{+}b_{-}\mathcal{N}_{k,l}^{+}+b_{+}b_{-}\mathcal{M}_{k,l}^{-}\right\} ,
\end{align*}
where

\[
\begin{array}{cl}
\mathcal{M}_{k,l}^{\pm}= & \frac{\left(\varrho_{k,l}(++)+\varrho_{k,l}(--)\right)}{2}\pm\frac{2w_{+-}\varrho_{k,l}(+-)}{Q}\\
 & \pm\frac{\left(\varrho_{k,l}(++)-\varrho_{k,l}(--)\right)\left(w_{++}-w_{--}\right)}{2Q},
\end{array}
\]

\[
\begin{array}{cl}
\mathcal{N}_{k,l}^{\pm}= & \frac{\left[\mp(\varrho_{k,l}(++)-\varrho_{k,l}(--))w_{+-}\pm\varrho_{k,l}(--)(w_{++}-w_{--})\right]}{2Qw_{+-}}\times\\
 & \left(w_{++}-w_{--}\pm Q\right).
\end{array}
\]

\textcolor{black}{All elements of average reduced density operator
of the dimer immerse between two impurities of the diamond structure
can be written as}

\begin{equation}
\rho=\left[\begin{array}{cccc}
\rho_{1,1} & 0 & 0 & 0\\
0 & \rho_{2,2} & \rho_{2,3} & 0\\
0 & \rho_{3,2} & \rho_{3,3} & 0\\
0 & 0 & 0 & \rho_{4,4}
\end{array}\right].\label{eq:rho-mat}
\end{equation}

Next, let us calculate the thermal entanglement behaviour of the dimer
Heisenberg isolated between two impurities.

\section{Thermal entanglement of two-qubit Heisenberg between two impurities}

In order to quantifying the thermal entanglement of the
Heisenberg qubits isolated in middle of the two impurities embedded
in the Ising-$XXZ$ diamond chain (see Fig. \ref{fig:diamond}), we
use the concurrence $\mathcal{C}$ define by Wootters \cite{woo}.

\begin{equation}
\mathcal{C}(\rho)=\mathrm{max}\{0,\sqrt{\lambda_{1}}-\sqrt{\lambda_{2}}-\sqrt{\lambda_{3}}-\sqrt{\lambda_{4}}\},\label{eq:Cdf-1}
\end{equation}
where $\lambda_{i}$ are eigenvalues in decreasing order of matrix
$R$,

\begin{equation}
R=\rho\cdot\left(\sigma^{y}\otimes\sigma^{y}\right)\cdot\rho^{*}\cdot\left(\sigma^{y}\otimes\sigma^{y}\right),
\end{equation}
which is constructed as a function of the density operator $\rho$
given by Eq. (\ref{eq:rho-mat}), with the asterisk denoting the complex
conjugate of matrix $\rho$.

Thereafter, \textcolor{black}{after a short algebra,} the corresponding
concurrence for our system can be reduced to

\begin{equation}
\mathcal{C}(\rho)=2\mathrm{max}\{0,|\rho_{2,3}|-\sqrt{\rho_{1,1}\rho_{4,4}}\}.\label{eq:conc-def}
\end{equation}

\begin{figure}
\includegraphics[scale=0.35]{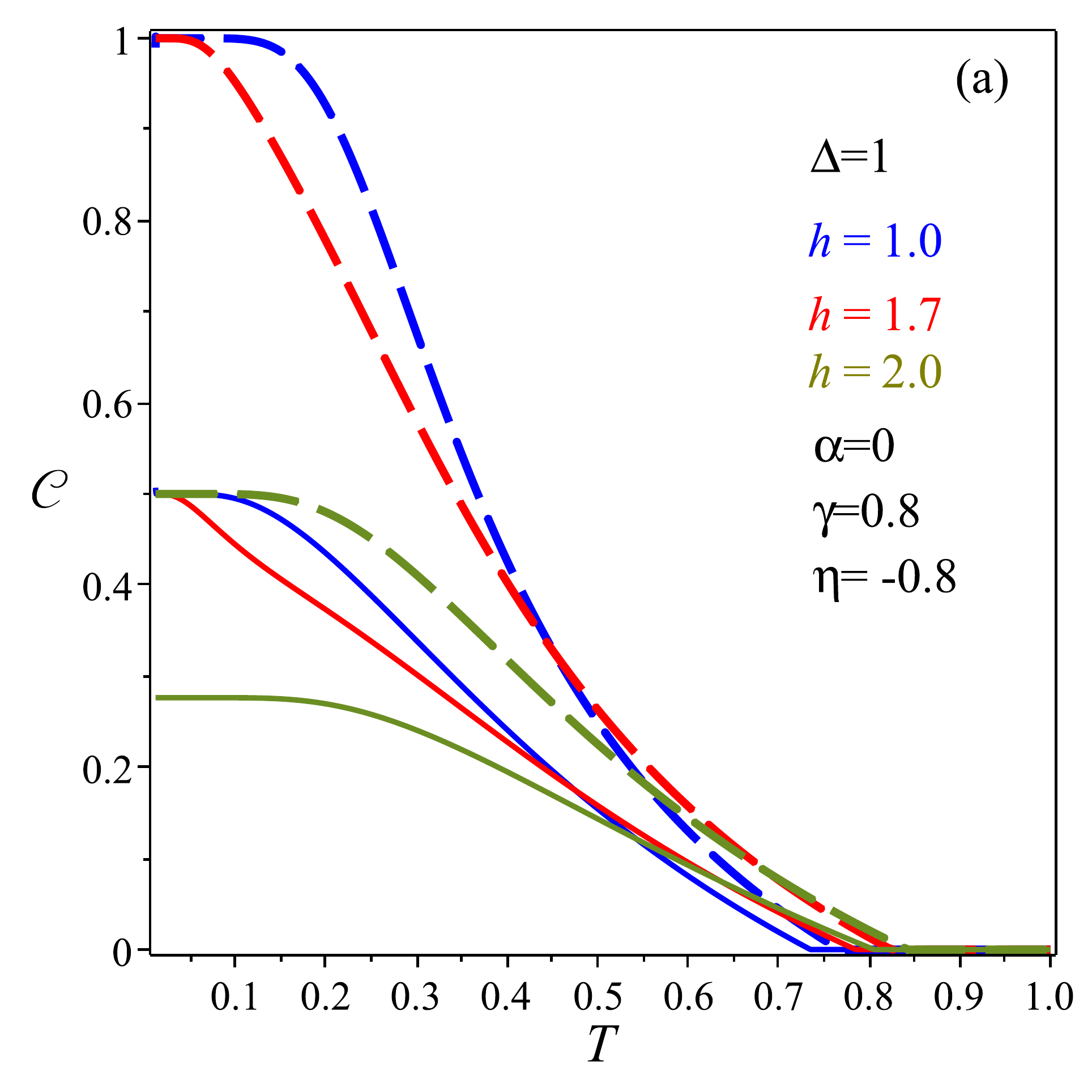}

\includegraphics[scale=0.35]{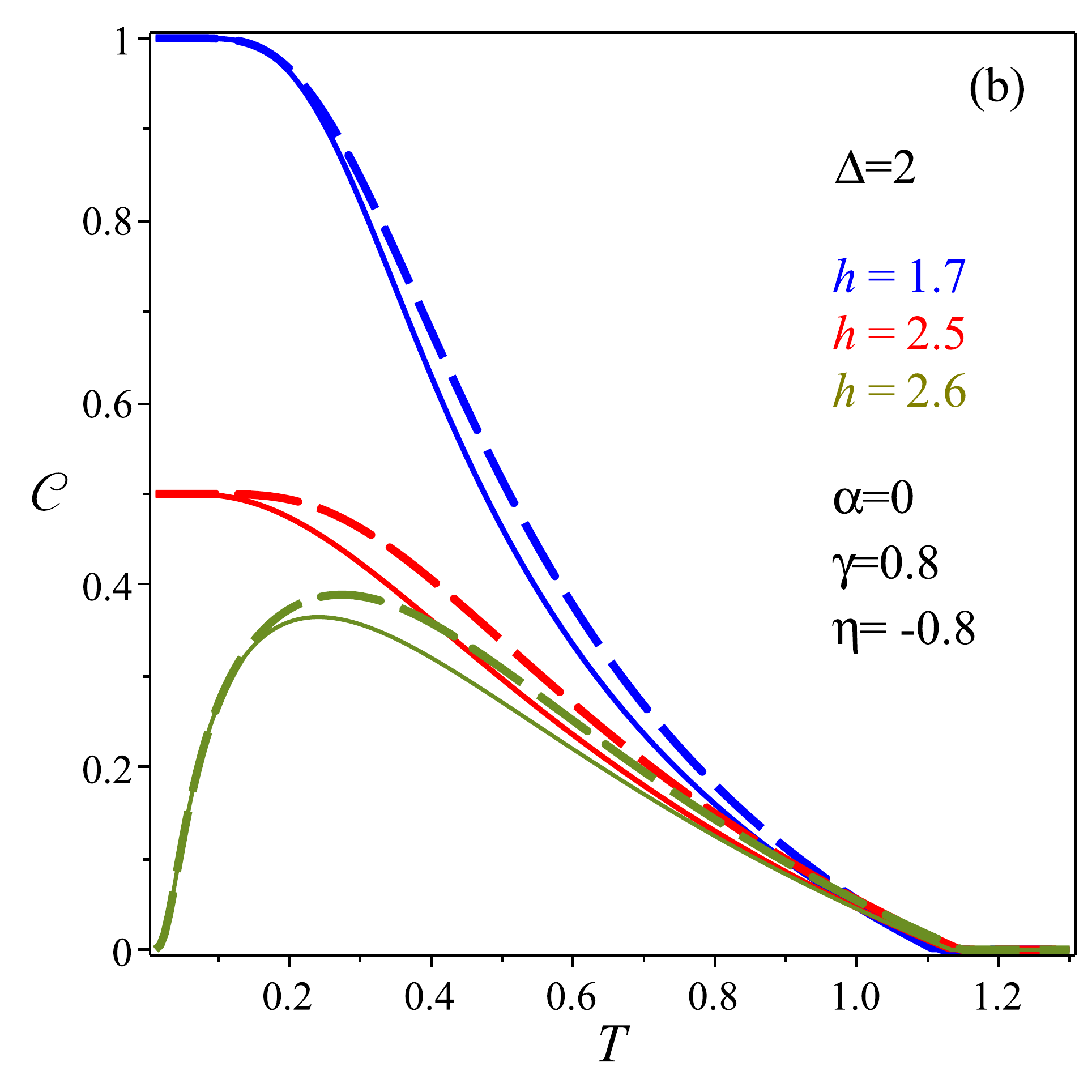}

\caption{\label{fig:C-Te}The concurrence as a function of temperature $T$
for different values of the magnetic field, with $J_{1}/J=1$. In
this figure we consider the original model without impurities (solid
line). On the other hand, for the model with impurities (dashed line),
we consider the parameters $\alpha=0$, $\gamma=0.8$, $\eta=-0.8$.
(a) $\Delta=1.0$; (b) $\Delta=2$. }
\end{figure}

The thermal entanglement properties for Ising-$XXZ$ diamond chain
structure have been investigated in Reference \cite{moi}. Here we
are mainly concerned about the effects on the thermal entanglement
caused by the impurities in the Ising-$XXZ$ diamond chain. More specifically,
let us focus on the thermal entanglement of the isolated two-qubit
Heisenberg (dimer) located between the impurities.

From now on, we will plot the curves of the original model with solid
lines, while for the Ising-$XXZ$ diamond chain with two impurities,
we will use dashed lines.

In Fig. \ref{fig:C-Te} we describe the behavior of the concurrence
$\mathcal{C}$ with respect of the temperature $T$ for different
values of the magnetic field $h$ and impurities parameters $\alpha=0$,
$\gamma=0.8$ and $\eta=-0.8$. In Fig. \ref{fig:C-Te}(a), we observed
a dramatic increase in the concurrence when the model has impurities.
From this figure, it can be clearly seen that, for the Ising-$XXZ$
diamond chain with two anisotropic impurities $(\widetilde{\Delta}=1.8)$,
the concurrence increases strongly, reaching the maximum entanglement
(see dashed line) for different values of the magnetic field. On the
other hand, in Fig. \ref{fig:C-Te}(b) we can see that for \textcolor{black}{high}
values of the anisotropy ($\Delta=2$ and $\widetilde{\Delta}=2.8$)
the effect of the impurities are weak, whereas for the temperatures
$T=0$ and $T\apprge1.1$ the effects of impurities are null.

\begin{figure}
 \includegraphics[scale=0.22]{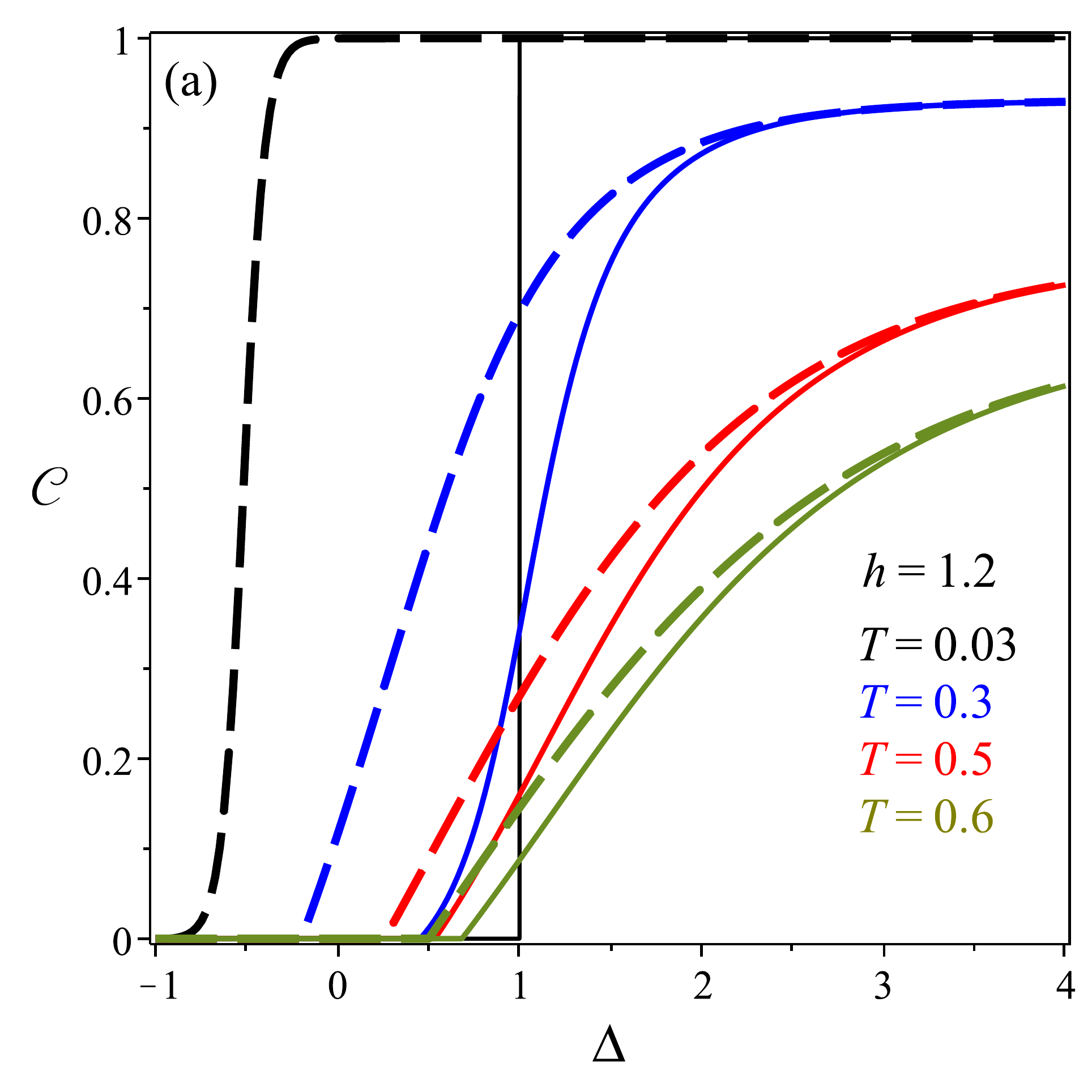}\includegraphics[scale=0.22]{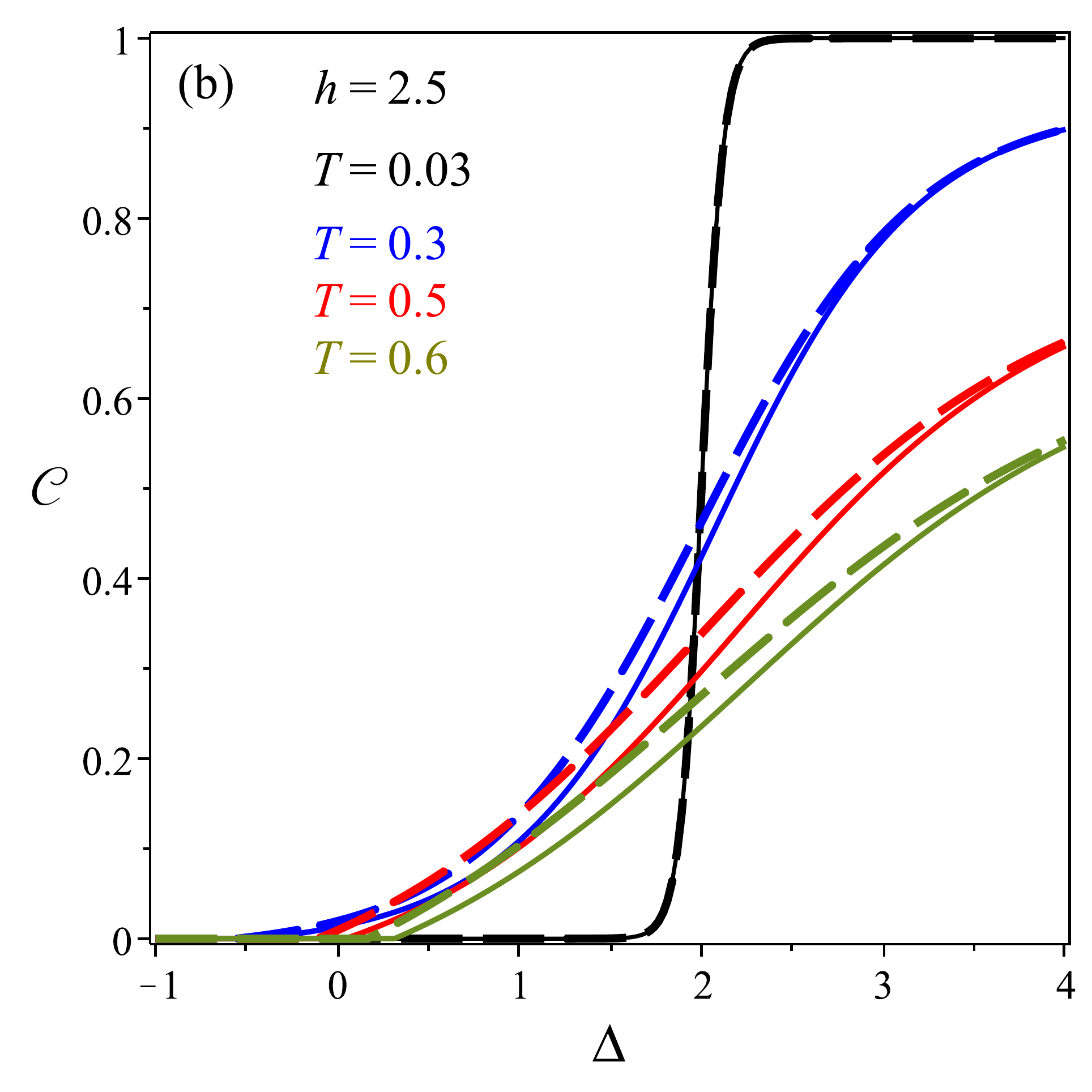}\caption{\label{fig:C-Delta} (Color online) The concurrence as a function
of the anisotropy factor $\Delta$ for different values of the temperature
$T$, with $J_{1}/J=1$. For the Ising-$XXZ$ model without impurities
we have $\alpha=\gamma=\eta=0$ (solid line). On the other hand, for
the model with impurities, we fixed $\alpha=0$, $\gamma=0.8$ and
$\eta=-0.8$ (dashed line). (a) $h=1.2$; (b) $h=2.5$.}
\end{figure}

In Fig. \ref{fig:C-Delta}, the concurrence $\mathcal{C}$ is plotted
as a function of anisotropy parameter $\Delta$ for different values
of the temperature and the fixed values of impurities parameters $\alpha=0$,
$\gamma=0.8$ and $\eta=-0.8$. \textcolor{black}{We can observe in
these figures that, for}\textcolor{red}{{} ${\color{black}h=1.2}$ }\textcolor{black}{and}\textcolor{red}{{}
${\color{black}h=2.5}$ }\textcolor{black}{we have transition from
unentangled ferrimagnetic (}\textcolor{red}{${\color{black}UFI}$}\textcolor{black}{)
to entangled state (}\textcolor{red}{${\color{black}ENQ}$}\textcolor{black}{)
and}\textcolor{red}{{} }\textcolor{black}{unentagled ferromagnetic state
(}\textcolor{red}{${\color{black}UFM}$}\textcolor{black}{) to same
entanglement state (}\textcolor{red}{${\color{black}ENQ}$}\textcolor{black}{)
respectively (see \cite{moi}).} As one can see from Fig. \ref{fig:C-Delta}(a)
the anisotropic behavior of the thermal concurrence for impurities
is more robust. We can also observe that the presence of impurities
will allow the existence of thermal entanglement in regions beyond
the reach of the original model. In Fig. \ref{fig:C-Delta}(b), we
can see that, \textcolor{black}{for} strong magnetic fields and low
temperatures, the effect of impurity disappears \textcolor{black}{are
gradually lost. Now,} when we increase the temperature, the thermal
entanglement it is slightly more robust in the presence of impurities.\textcolor{red}{{}
}\textcolor{black}{We can conclude that} the entanglement for the
model with impurities is much more intense than the model without
impurities in a broader range of the anisotropy parameter. 
\begin{figure}
\includegraphics[scale=0.35]{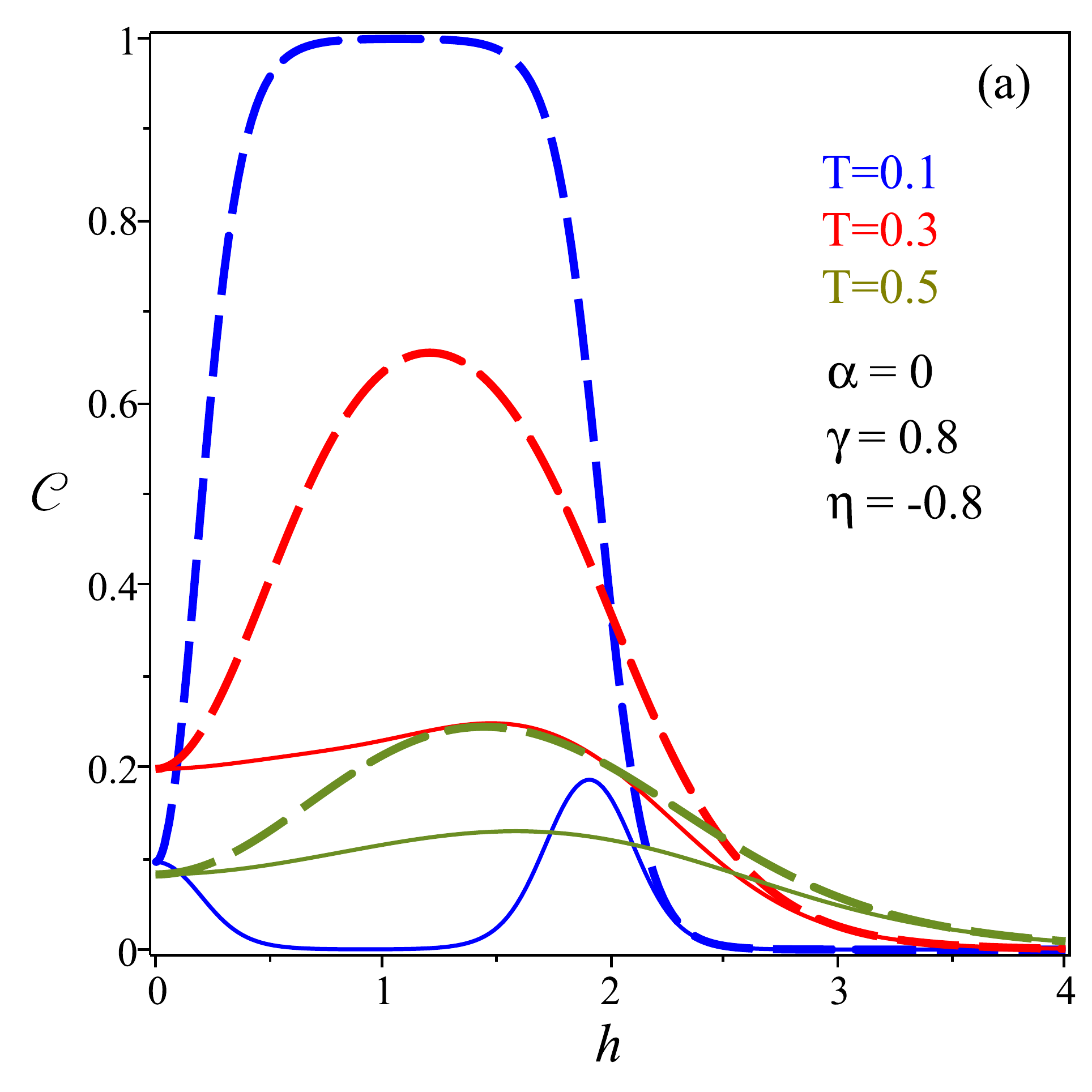}

\includegraphics[scale=0.35]{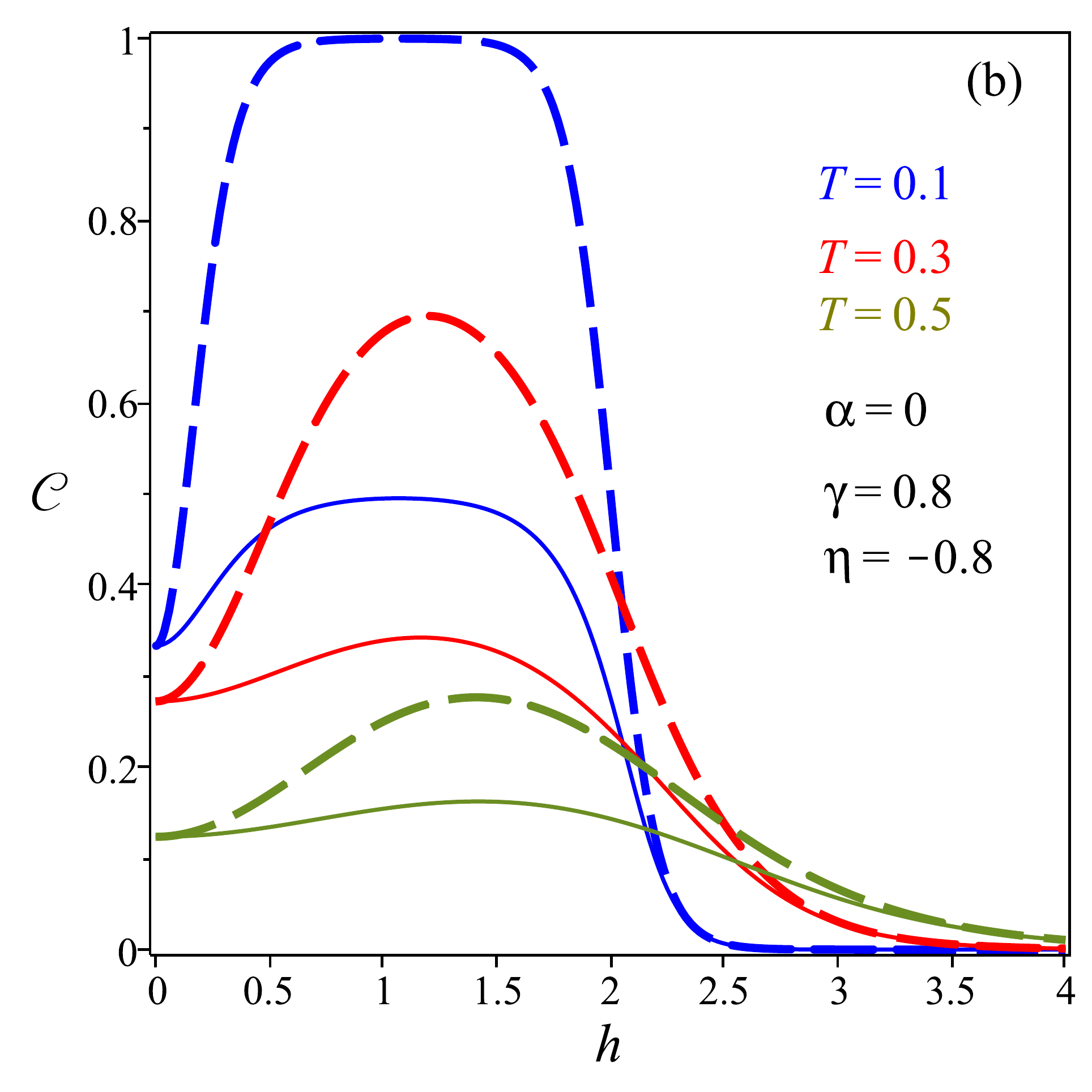}\caption{\label{fig:hvsC} (Color online) The concurrence as a function of
the magnetic field $h$, for $J_{1}/J=1$ and different values of
the temperature $T$. (a) $\Delta=0.9$; (b) $\Delta=1$.}
\end{figure}

In Fig. \ref{fig:hvsC}, the variations of concurrence $\mathcal{C}$
with magnetic field $h$ for different values of temperature $T$
and fixed values of the impurities parameter $\alpha=0$, $\gamma=0.8$
and $\eta=-0.8$ are plotted in two different cases (i.e $\Delta=0.9$
and $\Delta=1$). In Fig. \ref{fig:hvsC} (a), the curve for $T=0.1$
on original model becomes unentangled in the interval $0.6\apprle h\apprle1.4.$
In contrast, in our model with impurities this is maximally entangled,
signaling the strong influence of impurities in the entanglement of
the our model. Similar behavior has the concurrence when the temperature
increases. Thus, from \textcolor{black}{this}\textcolor{red}{{} }figure,
we can see \textcolor{black}{that} for $T=0.3$, the concurrence reaches
the value $\mathcal{C}\mathit{\approx\mathrm{0.24}}$ for the original
model, while for model with impurities we obtain $\mathcal{C}\mathit{\approx}0.65$.
While, for $T=0.5$, we obtain $\mathcal{C}\mathit{\approx}0.13$
in the original model, already in the model with impurities we obtained
$\mathcal{C}\mathit{\approx\mathrm{0.24}}$. In Fig. \ref{fig:hvsC}
(b) we can see the enhancing the concurrence \textcolor{black}{(dashed
line)} when compared to the model without impurities \textcolor{black}{(solid
line). Thus}, in low temperature $T=0.1$, we have a sudden increase
the concurrence until $\mathcal{C}\mathit{=\mathrm{1}}$ for the model
with impurities. However, for the model without impurities we obtained
$\mathcal{C}\mathit{=\mathrm{0.5}}$. Moreover, it can be observed
in this figure that, for higher temperatures (e.g $T=0.3$, $T=0.5$),
the entanglement is still greater than in the case without impurities.
We also observed that, for intense magnetic fields, the effect of
the impurities no longer influences in entanglement. 

\begin{figure}
\includegraphics[scale=0.23]{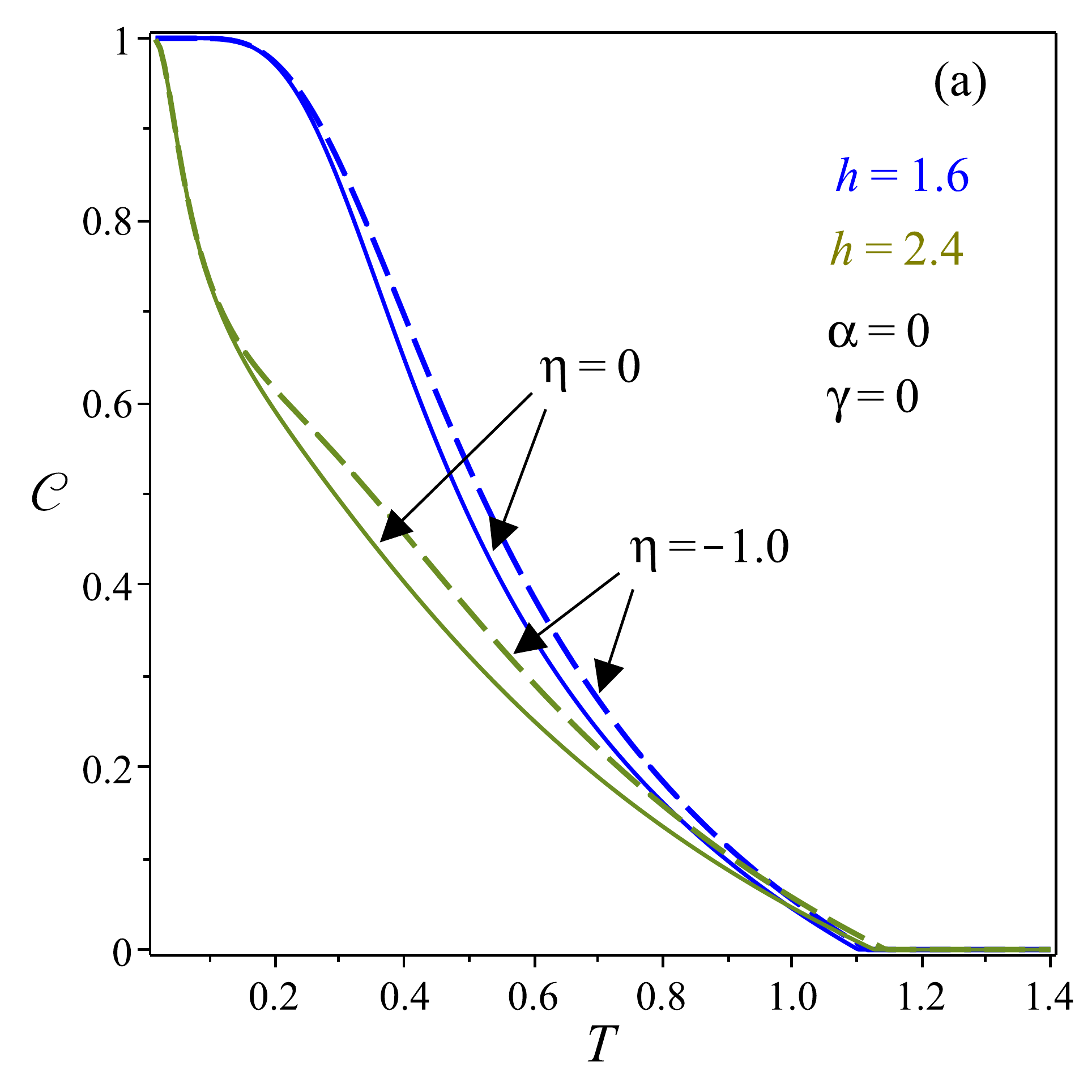}\includegraphics[scale=0.23]{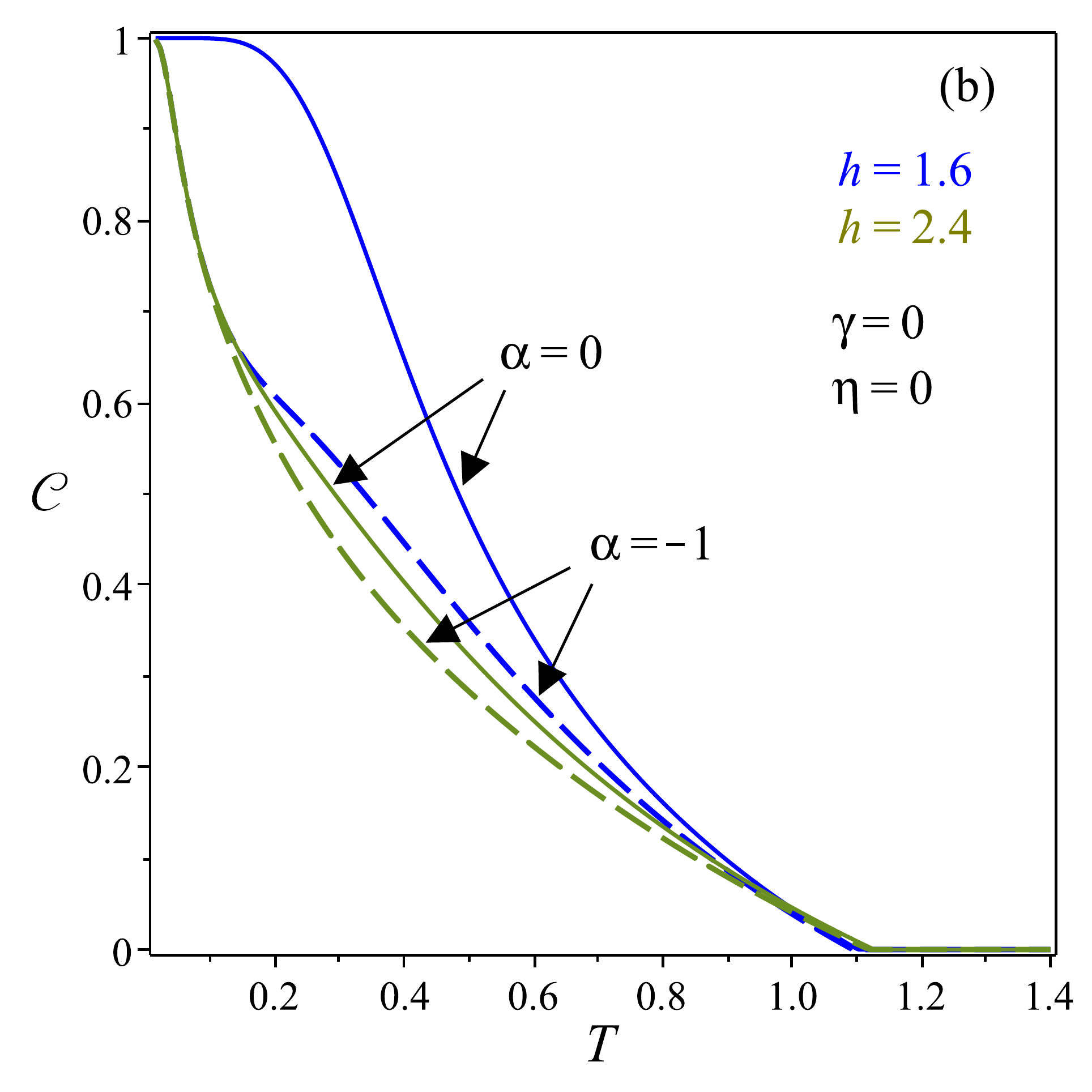}\caption{\label{fig:TvsC}(Color online) The concurrence versus temperature
for different values of the magnetic field $h=1.6$ (blue), $h=2.4$
(green), $J=J_{1}=1$, $\Delta=2$. The concurrence of the model with
impurity is show in dashed line. (a) $\eta=-1$ (dashed line); (b)
$\alpha=-1$ (dashed line).}
\end{figure}

Finally, in Fig. \ref{fig:TvsC} let us study the behavior of the
system in two extreme cases. In Fig. \ref{fig:TvsC} (a), we analyze
the case \textcolor{black}{when} $\widetilde{J}_{1}=0$ or $\eta=-1$
which corresponds to the collapse of Ising-like interaction of the
impurities with the dimer isolated from diamond chain. We observed
that we can enhanced the concurrence of the dimer located between
the impurities when it is isolated from the diamond chain. Is notable
that the entanglement can be changed by the presence of impurities
even in the case where the physical coupling (Ising type) to the impurity
is \textcolor{black}{null}. On the other hand, in Fig. \ref{fig:TvsC}
(b) let us consider $\widetilde{J}=0$ or $\alpha=-1$, this case
corresponds to the situation where the Heisenberg interaction in the
impurities is null. It easily can be seen that the thermal concurrence
decreases in the dimer located between the impurities. 

\section{Conclusions}

\textcolor{black}{In summary, in this paper we investigated the effects
of the impurities local sites on the Ising-}\textcolor{red}{${\color{black}XXZ}$
}\textcolor{black}{diamond chain}. In the model we have two impurities
with an isolated $XXZ$ dimer between them. In particular, we have
developed a method to solve exactly the Ising-$XXZ$ diamond chain
with two impurities. Our attention has been focused on the study the
thermal entanglement of the isolated $XXZ$ dimer. We demonstrate,
that the thermal entanglement can be controlled and tuned by introducing
impurities \textcolor{black}{into an Ising-$XXZ$ chain}.

As a matter of fact, for certain parameters of the impurities, the
thermal entanglement is maximum in low temperatures, unlike what happens
in the model without impurities. \textcolor{black}{On the other hand},
for high temperatures, the effects of the impurities is null. \textcolor{black}{Moreover},
the concurrence have been examined in detail with respect to the anisotropy
parameter, \textcolor{black}{we observed that impurities strongly}\textcolor{red}{{}
}\textcolor{black}{influence at the behavior of the entanglement at
low temperatures for weak magnetic fields}. \textcolor{black}{More
specifically, the results showns the existence of thermal entanglement
in regions}\textcolor{red}{{} }\textcolor{black}{beyond the reach of
the original model.}\textcolor{red}{{} }While, for strong magnetic fields
the effect of anisotropy on entanglement is very weak. Similar behavior
is obtained for entanglement as a function of the magnetic field.
For weak magnetic fields, the model is maximally entanglement in contrast
to the same model without impurities. \textcolor{black}{However, for
strong magnetic fields, impurities have no}\textcolor{red}{{} }\textcolor{black}{more
influence in the thermal entanglement for both models.} Furthermore,
\textcolor{black}{our results} sheds some light on the local \textcolor{black}{control}
of thermal entanglement by manipulating the parameters of the impurities
inserted in the Ising-$XXZ$ diamond chain. Finally, it is expected
\textcolor{black}{that this new approach may be helpful for realising
the quantum teleportation process of information.}

\section*{Acknowledgment}

O. Rojas, M. Rojas and S. M. de Souza thank CNPq, Capes and FAPEMIG
for partial financial support. I. M. Carvalho acknowledges Capes for
full financial support.

\end{document}